\documentclass[prb,twocolumn,notitlepage,longbibliography]{revtex4-2}
\usepackage{amsmath}
\usepackage{amssymb}

\usepackage[unicode=true,colorlinks=true,citecolor=blue,urlcolor=blue]{hyperref}

\usepackage{bm}
\usepackage{epsfig}

\usepackage[normalem]{ulem}

\renewcommand {\Im}{\mathop\mathrm{Im}\nolimits}

\renewcommand {\i}{{\rm i}}
\renewcommand {\phi}{{\varphi}}
\newcommand {\rmi}{{\rm i}}
\newcommand {\rmd}{{\rm d}}

\newcommand {\e}{{\rm e}}

\usepackage{physics}
\usepackage{ulem}

\begin{document}
\title{Ratchet effect in frequency-modulated waveguide-coupled emitter arrays
}

\author{Alexander N. Poddubny }
\email{poddubny@coherent.ioffe.ru}

\author{Leonid E. Golub}
\affiliation{Ioffe Institute, St. Petersburg 194021, Russia}

\begin{abstract}
We study theoretically the spatial distribution of the polarizations in the array of resonant electromagnetic dipole emitters coupled to a one-dimensional waveguide. The ratchet effect manifests itself in the spatial asymmetry of the distribution of the emitter occupations  along the array under  symmetrical  pumping from both  sides. The  occupation asymmetry  is driven  by the periodic modulation  in time of the emitter resonance frequencies. We find numerically and analytically the optimal conditions for maximal asymmetry. We also demonstrate that the ratchet effect can be enhanced due to the formation of topological electromagnetic edge states,  enabled by the  frequency modulation. Our results apply  to the classical structures with coupled resonators or arrays of semiconductor quantum wells as well as the quantum setups with waveguide-coupled natural or artificial atoms. 
\end{abstract}

\maketitle
\section{Introduction}\label{sec:intro}
Ratchets are periodic systems that generate a directed particle flow under an action of a time-oscillating force with zero mean. Ratchets are known in different fields of physics, chemistry and biology including Brownian motors, temperature ratchets as well as rocking and pulsating ratchets~\cite{Brownian_motors,EL_SG_2011,book_brownian_ratchets,Gulyaev2020}. Ratchets can be realized in semiconductor, cold atom, superconducting and active matter systems~\cite{Reichhardt}. Application of a static magnetic field
broadens this phenomenon allowing introducing magnetic ratchets~\cite{magn_1,magn_2}. In finite systems, the ratchet effect consists in the appearance of an inhomogeneous distribution of particle density in space-periodic systems under an action of a time-periodic driving.

\begin{figure}[t]
\centering\includegraphics[width=0.45\textwidth]{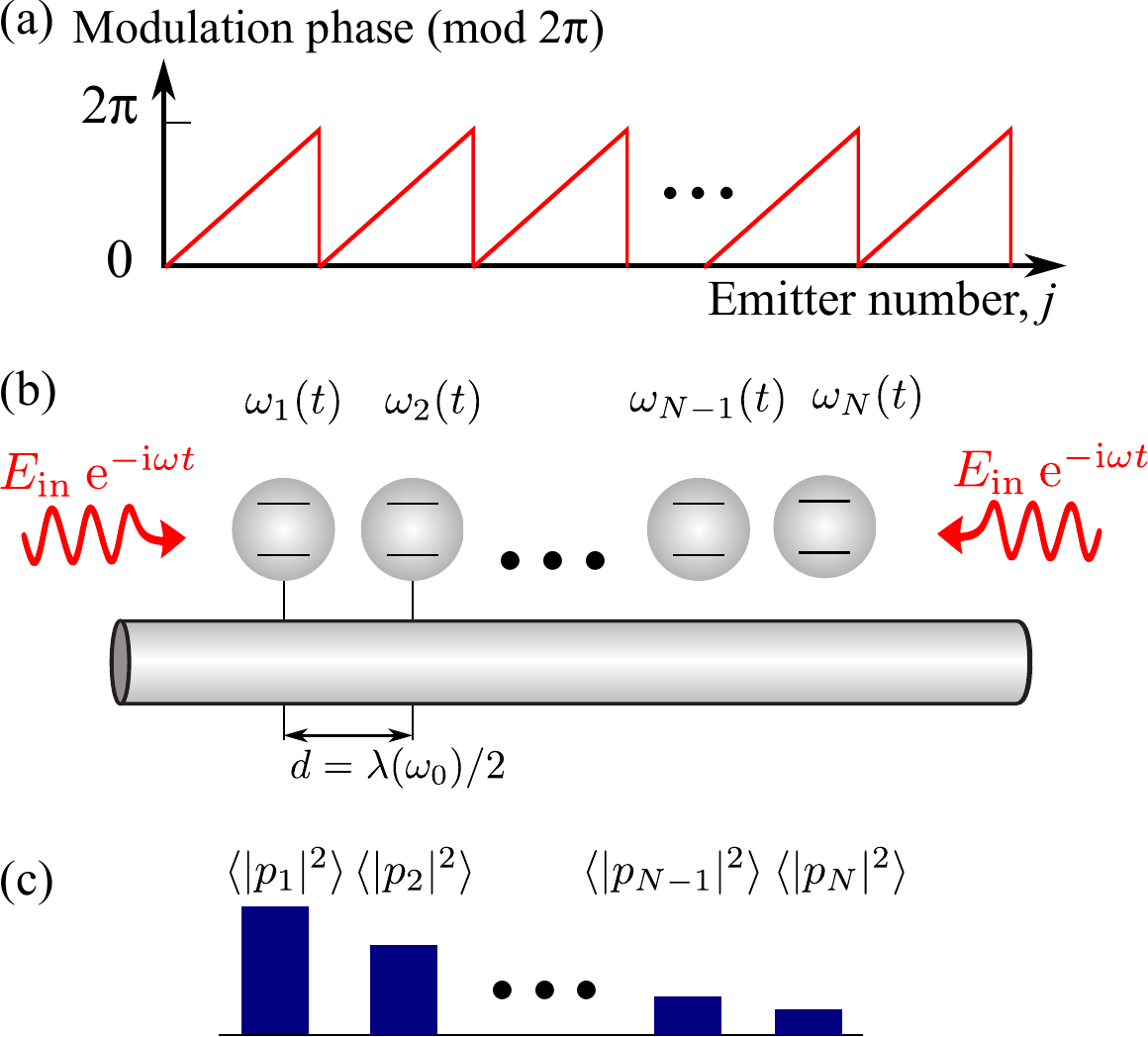}
\caption{ Schematics of array of emitters coupled to the waveguide with time-modulated resonant frequencies $\omega_j(t)$.  (a) Illustration of the asymmetric spatial dependence of the modulation phase.
(b)  Sketch of the structure  excited  from both  sides by a monochromatic wave with electric field $E_{\rm in}\e^{-\rmi \omega t}$. (c)
Spatially asymmetric  distribution of the time-averaged squared absolute values of the polarization $\left<|p_j|^2\right>$.
}\label{fig:1}
\end{figure}

In this work we theoretically study  the ratchet effect in an array of resonant electromagnetic emitters, coupled to a one-dimensional waveguide. The structure under consideration is schematically illustrated in Fig.~\ref{fig:1}. We consider a periodic array with a simple unit cell, that has an inversion symmetry and is symmetrically excited from both sides by an electromagnetic wave. We also introduce the time-dependent  modulation of the emitters resonance frequency. This modulation  breaks the spatial inversion symmetry  (Fig.~\ref{fig:1}a) and enables the ratchet effect, that is manifested in spatially asymmetric distribution of the occupations of the emitters (Fig.~\ref{fig:1}c). Such approach, based on frequency modulation, is well known in modern photonics and optomechanics~\cite{Estep2014,Fang2017,Chapman2017}, where it has been shown to break the time-reversal invariance and make the propagation of photons in the structure nonreciprocal, see the review~\cite{Sounas2017} for more details. Here, however, we focus on spatial asymmetry rather than nonreciprocity.  Our goal is to investigate  the asymmetry in the  distribution of the emitter polarizations depending on the parameters of the frequency modulation.

Our setup can be readily implemented in arrays of microwave resonators~\cite{Estep2014} or superconducting qubits \cite{Krantz2019} where the resonance frequency  can be controlled by an external electric current.  The waveguide-coupled qubit arrays are now actively studied in the context of  waveguide quantum electrodynamics~\cite{Roy2017,KimbleRMP2018,sheremet2021waveguide}.  Another potential realization  is offered by arrays of semiconductor quantum wells and quantum dots with excitonic optical resonances, where the exciton resonance frequency can be modulated by an acoustic wave~\cite{Akimov2015,Kuznetsov2020,Wigger2021}. 

The rest of the manuscript is organized as follows. Section~\ref{sec:model} outlines our theoretical model. Next, in Sec.~\ref{sec:2} we consider a simplest situation of just two frequency-modulated resonant emitters and present both numerical  and approximate analytical results. Section~\ref{sec:N} is devoted to the ratchet effect in longer emitter  arrays. We demonstrate, that the effect is enhanced due to the formation of topological edge states driven by the frequency modulation. Finally, Sec.~\ref{sec:summary} presents summary and outlook.

\section{Model}\label{sec:model}
We consider a one-dimensional periodically spaced array of emitters, schematically illustrated in Fig.~\ref{fig:1}. 
We assume, that the emitter resonance frequencies $\omega_j(t)$ are externally modulated in time as 
\begin{equation}\label{eq:omegaj}
\omega_j=\omega_0+2A\cos(\Omega t {-}\alpha_j)\:, \quad j=1\ldots N\:,
\end{equation}
where $\omega_0$ is the resonance frequency without modulation, and $A$ is the modulation amplitude that is for simplicity assumed to be identical for all the emitters. The emitters are assumed to be small as compared to the light wavelength at the frequency $\omega_0$ and their electromagnetic properties can be described in the dipole approximation by  electric dipole moments $p_j$.

The structure is excited symmetrically from both sides by an coherent electromagnetic wave, as shown in  Fig.~\ref{fig:1}.   The wave is polarized transverse to the waveguide. 
We are interested in the distribution of the emitter polarizations in the presence of the modulation Eq.~\eqref{eq:omegaj}. This distribution can be found from the following system of linear equations
\begin{multline}\label{eq:gen}
\left(\rmi\frac{\rmd}{\rmd t}-\omega_0-2A\cos(\Omega t {-} \alpha_j)+\rmi \gamma\right)p_j\\+\rmi \gamma_{\rm 1D}\sum\limits_{j'=1}^N 
\e^{\rmi\varphi|j-j'|}p_{j'}= E_j\e^{-\rmi \omega t}.
\end{multline}
Here, $\omega$ is the excitation frequency and $E_j$ is  the electric field amplitude of the incident waves at the $j$-th emitter up to the constant common factor $\propto \sqrt{\gamma_{\rm 1D}}$~\cite{Suh2004}. Equations~\eqref{eq:gen} present a generalization of the usual discrete dipole approximation~\cite{Draine1994} and mode coupling  theory~\cite{Suh2004} for a system, modulated in time. We also note, that while  Eqs.~\eqref{eq:gen} are essentially  classical and describe just an array of coupled resonant oscillators, the same system of equations can be applied to a setup of waveguide quantum electrodynamics, describing an array of cold atoms or superconducting qubits, coupled to a waveguide~\cite{Caneva2015,Molmer2019,Ke2019}. In the quantum case the polarizations $p_j$ correspond to the amplitudes of the excited states of the atoms or qubits. The equations~\eqref{eq:gen} are valid provided that the system is excited by a weak coherent wave, so that no more than one photon is present in the system.
 We have also added to Eqs.~\eqref{eq:gen} a phenomenological nonradiative damping $\gamma$ describing all other mechanisms of the polarization decay besides emission into the waveguide.

In order to solve Eqs.~\eqref{eq:gen} we expand the amplitudes in the Fourier series
\begin{equation}
p_j=\sum\limits_{m=-\infty}^\infty p_j^{(m)}\e^{-\rmi (\omega {-} m\Omega) t}
\end{equation}
which yields the following system of linear equations for the Fourier harmonics $p_j^{(m)}$:
\begin{multline}\label{eq:hmain}
\left(\omega-\omega_0 {-} m\Omega
+\rmi \gamma\right)p_j^{(m)}-A\qty(\e^{{-}\rmi \alpha_j}p_j^{(m-1)}+\e^{\rmi \alpha_j}p_j^{(m+1)})\\+\rmi \gamma_{\rm 1D}\sum\limits_{j'=1}^N 
\e^{\rmi\varphi|j-j'|}p_{j'}^{(m)}=E_j\delta_{m,0}.
\end{multline}
The system of Eqs.~\eqref{eq:hmain} can be readily solved numerically after the Fourier series are truncated. In what follows we are interested in the mirror asymmetry of the polarization distribution
\[\abs{p_j(t)}^2 = \sum\limits_{m=-\infty}^\infty \abs{p_j^{(m)}}^2\:,
\]
induced by the modulation.

{We note that the ratchet effect --- a space asymmetry of the polarization distribution --- is enabled by nonzero  damping. In the limit of zero damping, an additional time-inversion symmetry is present, and 
the system~\eqref{eq:hmain} where the substitution $\alpha_j \to -\alpha_j$ is made has the solutions that are complex-conjugated to those in the initial system. This complex conjugation does not affect $\abs{p_j}^2$ and, hence, the population distribution $|p_j|^2$ is a symmetric function of $j$.
}

Before proceeding to the discussion of the asymmetry it is instructive to consider  first a situation when coupling between the emitters in Eqs.~\eqref{eq:hmain} is neglected  and they are all modulated independently, i.e. the term 
$\e^{\rmi\varphi|j-j'|}$ is replaced by $\delta_{jj'}$. Physically, this situation corresponds to the case of very large nonradiative damping $\gamma$. The corresponding problem is known in literature and has been considered in the cavity optomechanics setup~\cite{Marquardt2006}. The solution is briefly presented below. The amplitude $p_j$ 
is given by 
\begin{equation}
p_j = -{\rm i}E_j \e^{-\rmi \Phi_j(t)} \int\limits_{-\infty}^t \dd t' \e^{-\rmi \omega t' + \rmi \Phi_j(t')},
\end{equation}
where $\Phi_j(t) = (\omega_0-\rmi \gamma)t+a\sin{\qty(\Omega t {-} \alpha_j)}$ with 
\begin{equation}
\label{a}
a={2A \over \Omega}.
\end{equation}
Expanding $\exp[\rmi \Phi_j(t')]$ by a series of Bessel functions, we perform integration and obtain 
\begin{equation}
\label{eq:amplitudes}
p_j = -E_j \e^{-\i\omega t- \i a\sin\qty(\Omega t {-} \alpha_j)} \sum_{n=-\infty}^\infty {J_{n}(a) \e^{\i n \qty(\Omega t {-} \alpha_j)}\over n \Omega -\Delta -\i\gamma_{\rm tot}}.
\end{equation}
Here we introduced {the detuning and the total damping}
\begin{equation}
\Delta = \omega-\omega_0, \qquad \gamma_{\rm tot}=\gamma+\gamma_{\rm 1D}.
\end{equation}
The time-averaged squared  absolute amplitudes have the form
\begin{equation}\label{eq:population}
\langle |p_j|^2\rangle = 
\abs{E_j}^2\sum_{n=-\infty}^\infty {J_{n}^2(a) \over (\Delta-n\Omega)^2 +\gamma_{\rm tot}^2},
\end{equation}
where angular brackets denote the averaging over time.
We note, that in this regime we still retain the full resonance width  $\gamma_{\rm tot}=\gamma+\gamma_{\rm 1D}$  in Eq.~\eqref{eq:population} and do not neglect $\gamma_{\rm 1D}$ as compared to $\gamma$. The reason behind this is that in state-of-the art emitter arrays the radiative decay can be made  much larger than the nonradiative one, $\gamma\ll \gamma_{\rm 1D}$~\cite{brehm2020waveguide,Mirhosseini2019}.

Equation \eqref{eq:population} demonstrates that the modulation leads to appearance of multiple resonance peaks at the frequencies 
$\omega_0+n\Omega$, shifted from the original  resonance position, that are analogous to Stokes and anti-Stokes resonances in the Raman scattering problem. The peaks amplitudes are controlled by  the values of the corresponding  Bessel functions $J_{n}^2(a)$. It can be shown following the Bessel function properties that the total number of peaks resolved in the spectrum is on the order  of 
$a$, Eq.~\eqref{a},
i.e. it increases linearly with the modulation amplitude.

\section{Asymmetry for a pair of modulated  emitters}\label{sec:2}
We will now describe the modulation-induced asymmetry for the simplest possible array with just two emitters. From now on we consider the arrays with a quarter-wavelength spacing, when $\varphi=\pi/2$. In this case  the coupling between the emitters is non-dissipative, $\Im [\rmi \exp(\rmi\varphi|m-n|)]=0$~\cite{vladimirova1998ru,vanLoo2013}. 
\begin{figure}[b]
\centering\includegraphics[width=\linewidth]{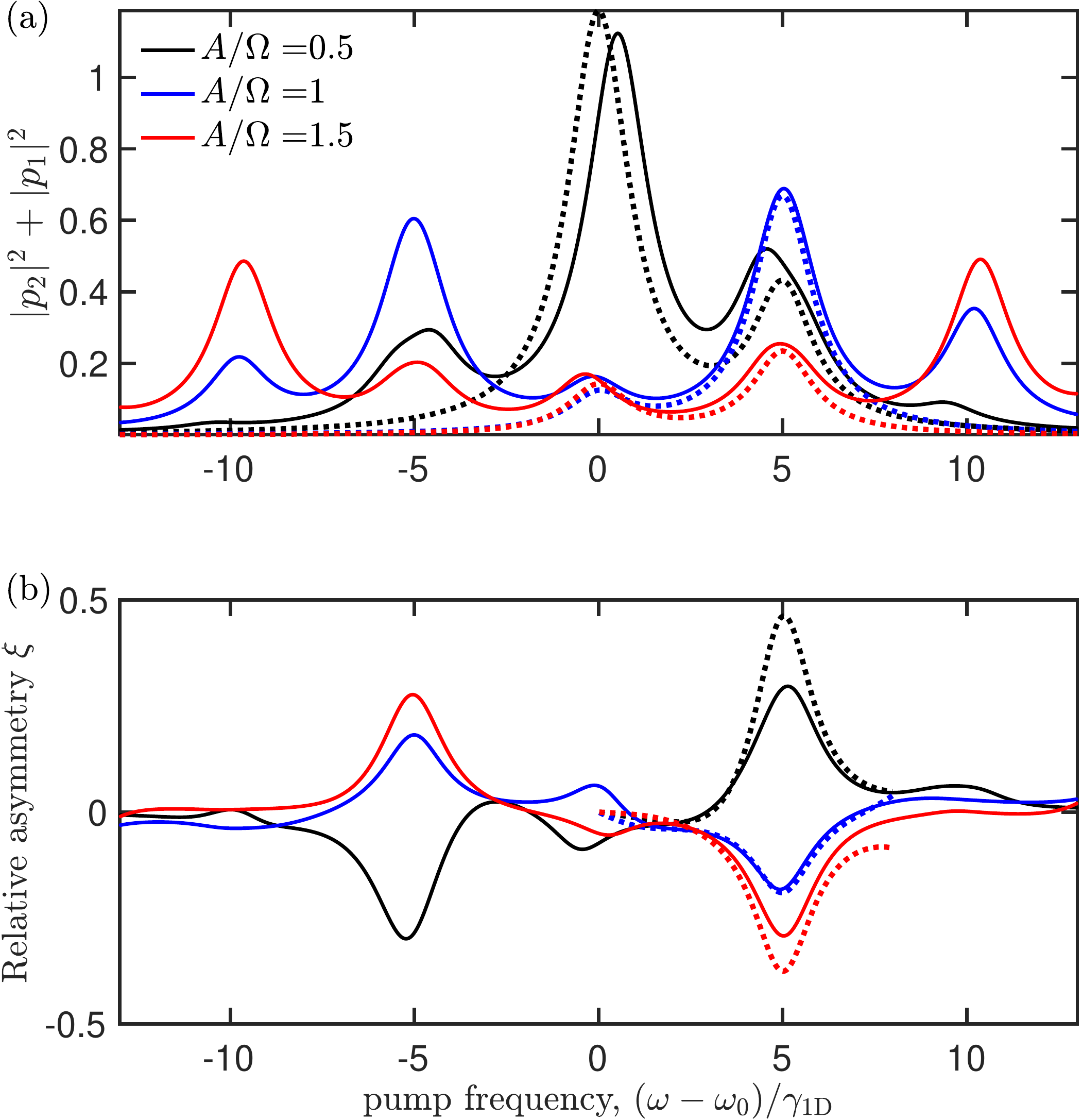}
\caption{Occupation (a) and occupation asymmetry (b) of two emitters calculated for three different modulation strength indicated on graph. 
Solid lines correspond to results of numerical calculation, dotted curves show analytical results of Eq.~\eqref{eq:population} and 
 Eq.~\eqref{eq:asymmetry}, respectively.
Calculation has been performed for $\mathcal E=1$, $\gamma=0$, $\Omega/\gamma_{\rm 1D}=5$. 
}\label{fig:2}
\end{figure}
Figure~\ref{fig:2} presents the results of numerical calculation of the total occupation of two emitters $\left< |p_1|^2+|p_2|^2 \right>$ [panel (a)] and the relative  asymmetry $\xi=\left<|p_{2}|^2-|p_{1}|^2\right>/(2\left<|p_1|^2+|p_2|^2\right>)$ [panel (b)] depending on the excitation frequency $\omega$. The calculation has been performed for three amplitudes of the modulation. 
The phases of modulation were $\alpha_1=-\pi/4$, $\alpha_2=+\pi/4$. In agreement with the analytical equation Eq.~\eqref{eq:population} the spectra in Figure~\ref{fig:2}(a) consist of  Lorentzian peaks at  the frequencies $\omega_0\pm \Omega$, $\omega_0\pm 2\Omega\ldots$. The amplitude of higher order peaks  increases with the modulation amplitude $A$. Numerically calculated spectra are  in satisfactory agreement with the analytical calculation following Eq.~\eqref{eq:population} (dotted curves), and including only two harmonics with $m=0$ and $m=1$.
There exists however an interesting feature in the numerically obtained spectra that is not captured by the approximate  Eq.~\eqref{eq:population}: they do not have mirror symmetry with respect to the individual emitter resonance at $\omega=\omega_0$, see e.g. the black curve in  Fig.~\ref{fig:2}(a). The reason behind this is the spatial structure of the eigenmodes of the coupled  pair of emitters. Namely, the single-excited eigenstates of the system \eqref{eq:gen} without the modulation have $p_1=\pm p_2=\pm 1/\sqrt{2}$ with the eigenfrequencies $\omega=\omega_0\pm \gamma_{\rm 1D}$ and equal radiative decay rates. Symmetrical excitation probes only the even eigenstate at $\omega=\omega_0+ \gamma_{\rm 1D}$ which explains why the black spectrum in Fig.~\ref{fig:2}(a) is slightly blueshifted from $\omega_0$. Since the approximate equation \eqref{eq:population} does not take into account the coupling between the emitters, it can not describe this blueshift.

The behavior of the occupation asymmetry, shown in Fig.~\ref{fig:2}(b) is more subtle. While it also has  peaks at the 
 $\omega_0\pm \Omega$, $\omega_0\pm 2\Omega\ldots$, both the  signs and the magnitudes of the peak amplitude can nonmonotonously depend both on the harmonic number and on the modulation strength. For example, increase of the modulation from $A=0.5\Omega$ to $A=\Omega$ flips the asymmetry spectrum near the frequency $\omega_0+\Omega$, as can be seen from comparison of black and blue curves in  Fig.~\ref{fig:2}(b). 
 
 In order to provide more insight into these numerical results we  develop below an analytical  perturbation theory in the limit of weak coupling between the emitters, $\gamma_{\rm 1D} \ll A, \abs{\omega-\omega_0}$. At $\gamma_{\rm 1D}=0$, the polarizations $p_{1,2}$ are given by Eq.~\eqref{eq:amplitudes}. The corrections $\delta p_{1,2}$ are found from Eq.~\eqref{eq:gen} in the first order in $\gamma_{\rm 1D}$:
\begin{multline}
\delta p_{1,2} = \gamma_{\rm 1D}E_{2,1}\e^{-\i\omega t- \i a\sin\qty(\Omega t {\pm} {\alpha/2})}\\ \times
\sum_{k=-\infty}^\infty  \e^{\i k(\Omega t {\mp} {\alpha / 2})}
\sum_{n=-\infty}^\infty {J_{n-k}(a)J_{n}(a) \over n \Omega -\Delta -\i\gamma_{\rm tot}}
\\ \times
 \sum_{n'=-\infty}^\infty {J_{n'}(a) \e^{\i n' \qty(\Omega t {\pm} {\alpha / 2})}\over (n'+k) \Omega  -\Delta -\i\gamma_{\rm tot}}.
\end{multline}
Here we take $\alpha_{1,2}={\mp} \alpha/2$ and $\varphi =\pi/2$ (anti-Bragg condition).

We define the occupation asymmetry ${\Xi}$ as follows
\begin{equation}
{\Xi} = {\left< |p_2|^2-|p_1|^2 \right>.  }
\end{equation}
For symmetric pumping $E_1=E_2 = \mathcal{E}$ we obtain:
\begin{multline}
{\Xi}= 4\gamma_{\rm 1D}\gamma_{\rm tot}\mathcal{E}^2
\sum_{k,m=-\infty}^\infty {J_{m}(a) J_{m-k}(a) \sin{k \alpha}\over (\Delta-m\Omega)^2 +\gamma_{\rm tot}^2}\\ \times
\sum_{n=-\infty}^\infty {J_{n-k}(a)J_{n}(a) \over (\Delta-n \Omega)^2 + \gamma_{\rm tot}^2}.
\end{multline}
Using the summation theorem for Bessel functions we finally get
\begin{multline}
\label{eq:asymmetry}
{\Xi} = 4\gamma_{\rm 1D}\gamma_{\rm tot}\mathcal{E}^2 \!\sum_{m,n=-\infty}^\infty \!\sin\qty[{m-n\over 2}\pi \mathop\mathrm{sgn}{\alpha} +(m+n){\alpha\over 2}]\\\times{J_{m}(a)J_{n-m}\qty(2a\abs{\sin{\alpha\over 2}})J_{n}(a)\over [(\Delta-m\Omega)^2 +\gamma_{\rm tot}^2][(\Delta-n\Omega)^2 +\gamma_{\rm tot}^2]} .
\end{multline}
The occupation asymmetry ${\Xi}$ is an odd function of the phase difference $\alpha$, as expected.
The analytically calculated  asymmetry spectra ${\xi=\Xi/(2\left<|p_1|^2+|p_2|^2\right>)}$ following Eqs.~\eqref{eq:asymmetry},~\eqref{eq:population} {with $\alpha=\pi/2$}, are shown in Fig.~\ref{fig:2}(b) by the dotted curves and satisfactory describe our numerical results.

\begin{figure}[t]
\centering\includegraphics[width=\linewidth]{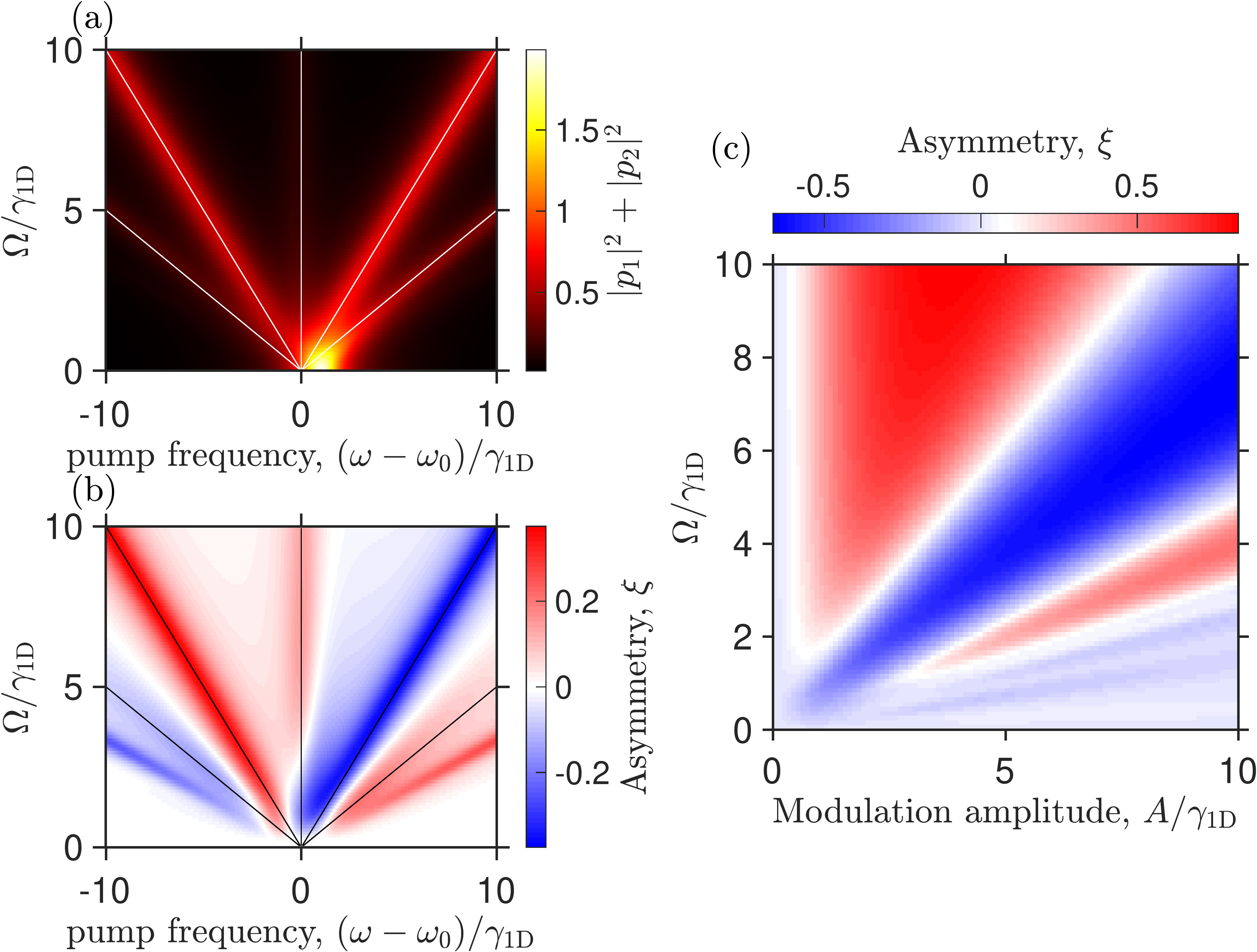}
\caption{(a,b) Color maps of emitter occupation (a) and  occupation asymmetry (b) depending on the pump and modulation frequencies for $A/\Omega=1$. Lines show positions of resonances $\omega-\omega_0=0\pm \Omega,\pm 2\omega$. (c) Dependence of asymmetry on modulation frequency and amplitude for $\omega-\omega_0=\Omega$. 
Calculation has been performed for $\mathcal E=1$, $\gamma=0$. }\label{fig:3}
\end{figure}

Figure~\ref{fig:3} examines in more detail how the spectra of the total occupation and the {relative} occupation asymmetry $\xi$ depend on the modulation frequency $\Omega$ and amplitude $A$. Panels (a) and (b) show the color maps of the occupation and asymmetry spectra depending on the modulation frequency. They demonstrate how a fan of resonances at $\omega_0\pm \Omega,\omega_0\pm 2\Omega$ appears in the spectrum. The calculation has been performed  for a fixed ratio $A/\Omega=1$. For this modulation strength the largest Bessel function $J_{n}(2A/\Omega)$ is the one at
$n=1$. Following Eqs.~\eqref{eq:population},\eqref{eq:asymmetry} this explains why the strongest features in Fig.~\ref{fig:3}(a,b) are  those at 
$\omega=\omega_0\pm\Omega$.

Figure~\ref{fig:3}(c) presents the color map of the  asymmetry parameter $\xi$ for a fixed frequency $\omega=\omega_0+\Omega$ depending on both $\Omega$ and $A$. The numerical calculation indicates that the strongest asymmetry at this frequency is  at $2A/\Omega\approx 1$.  For the chosen pumping frequency the asymmetry spectrum Eq.~\eqref{eq:asymmetry} is mainly  contributed by $m=n=1$. As such, the  asymmetry strength is determined by the Bessel function product $J_{1}^2(a)J_0(\sqrt{2}a)$. The maximum that is realized at $a\equiv 2A/\Omega\approx 1$, in agreement with the numerical results.

\section{Array of emitters}\label{sec:N}

\begin{figure}[b]
\centering\includegraphics[width=\linewidth]{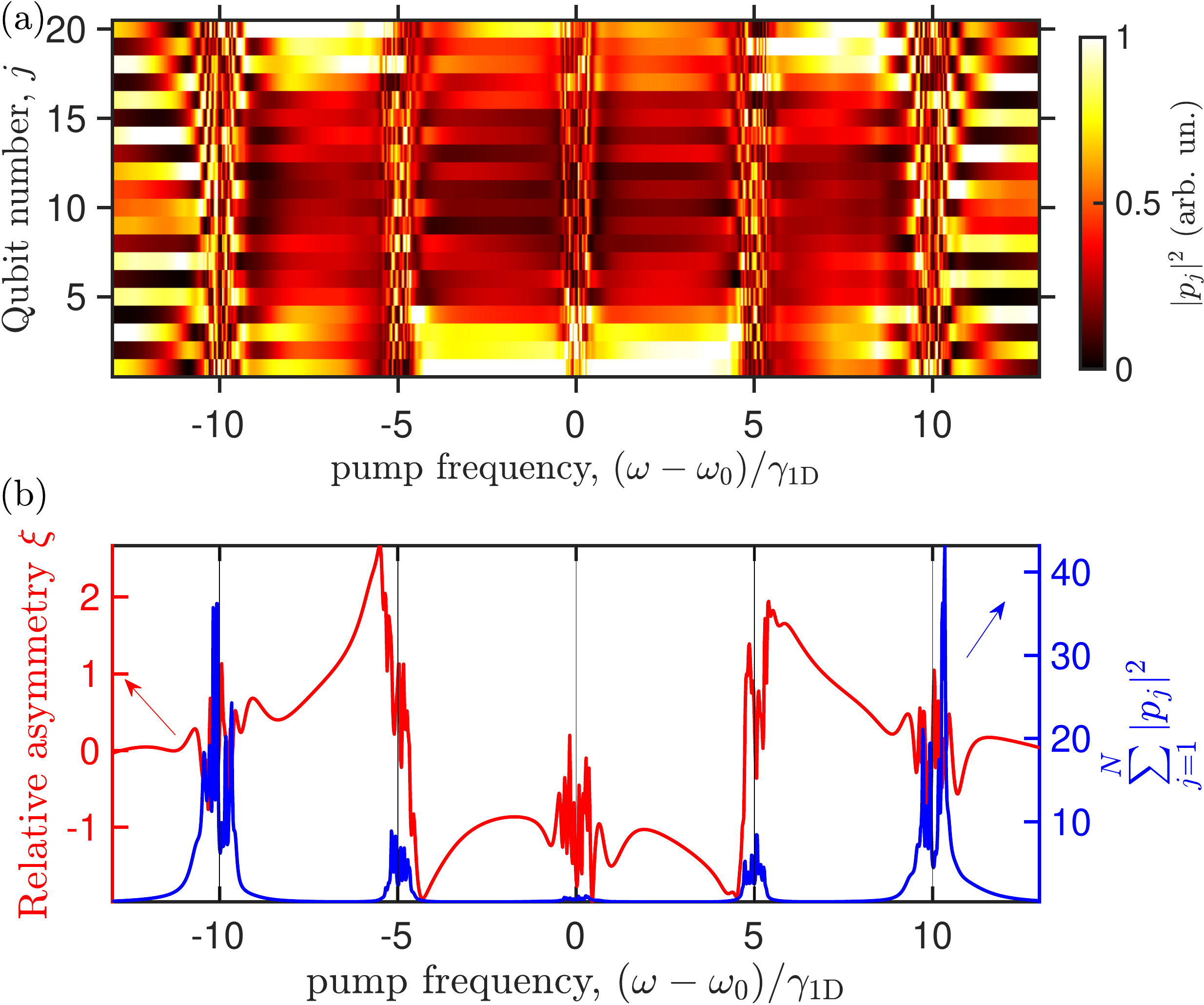}
\caption{(a) Color map of the occupation distribution $\left<|p_j(\omega)|^2\right>$ depending on the pump frequency and the  number of emitter $j$. Calculated for $N=20$, ${A}=\Omega=5\gamma_{\rm 1D}$, $\gamma=0$.
(b) Average first momentum of the occupation distribution 
${\xi}$
and total occupation number $\sum_j^N\left<|p_j|^2\right>$  in the array of $N=20$~emitters depending on the pump frequency. 
}\label{fig:4}
\end{figure}

We now proceed to the long arrays with $N\gg 1$ emitters. We consider the case, when the modulation phase is linearly changing along the array, $\alpha_j=j\pi/2$, see also  Fig.~\ref{fig:1}(a).
The results of numerical calculation are presented in Fig.~\ref{fig:4}. Figure~\ref{fig:4}(a) presents the dependence of the  occupation distribution in the array  $\left<|p_j(\omega)|^2\right>$ in the array on the emitter  number $j$ (vertical axis) and on the excitation frequency (horizontal axis) for $N= 20$ emitters. Figure~\ref{fig:4}(b)  shows the total   occupation $\sum_j\left<|p_j|^2\right>$ (blue curve, right $y$-axis) and the asymmetry parameter 
\begin{equation}\label{eq:x}
\xi=\frac{\sum_{j=1}^N{} (j-N/2-1/2)\left<|p_j(\omega)|^2\right>}{\sum_{j=1}^N\left<|p_j(\omega)|^2\right>}\:,
\end{equation}
that characterizes the deviation of the center of the occupation distribution from the array center at ${j=(N+1)/2}$ (red curve, left $y$-axis).
The calculation has been performed for a relatively large modulation frequency ${\Omega=5\gamma_{\rm 1D}}$ and the modulation amplitude $A=\Omega$. The main numerical results are as follows: (i) The occupation  distribution  has peaks at the Stokes and anti-Stokes frequencies $\omega_0\pm \Omega,\omega_0\pm 2\Omega$ and is strongly suppressed between these frequencies (see blue curve in Fig.~\ref{fig:4}(b)). (ii) In the intermediate spectral regions between Stokes and anti-Stokes frequencies, where the overall occupation is small, the  asymmetry parameter Eq.~\eqref{eq:x} is quite large, see red curve in Fig.~\ref{fig:4}(b). Moreover, the occupation distribution in the intermediate spectral regions,
$\omega_0\le \omega\le \omega_0+\Omega$ and  
$\omega_0+\Omega\le \omega\le \omega_0+2\Omega$ is not merely asymmetric but is concentrated at either left or right edge of the array as can be seen from the color maps in Fig.~\ref{fig:4}(a). Our numerical calculations indicate that such concentration of the occupation at the array edges becomes even more prominent for larger array lengths.

\begin{figure}[t]
\centering\includegraphics[width=\linewidth]{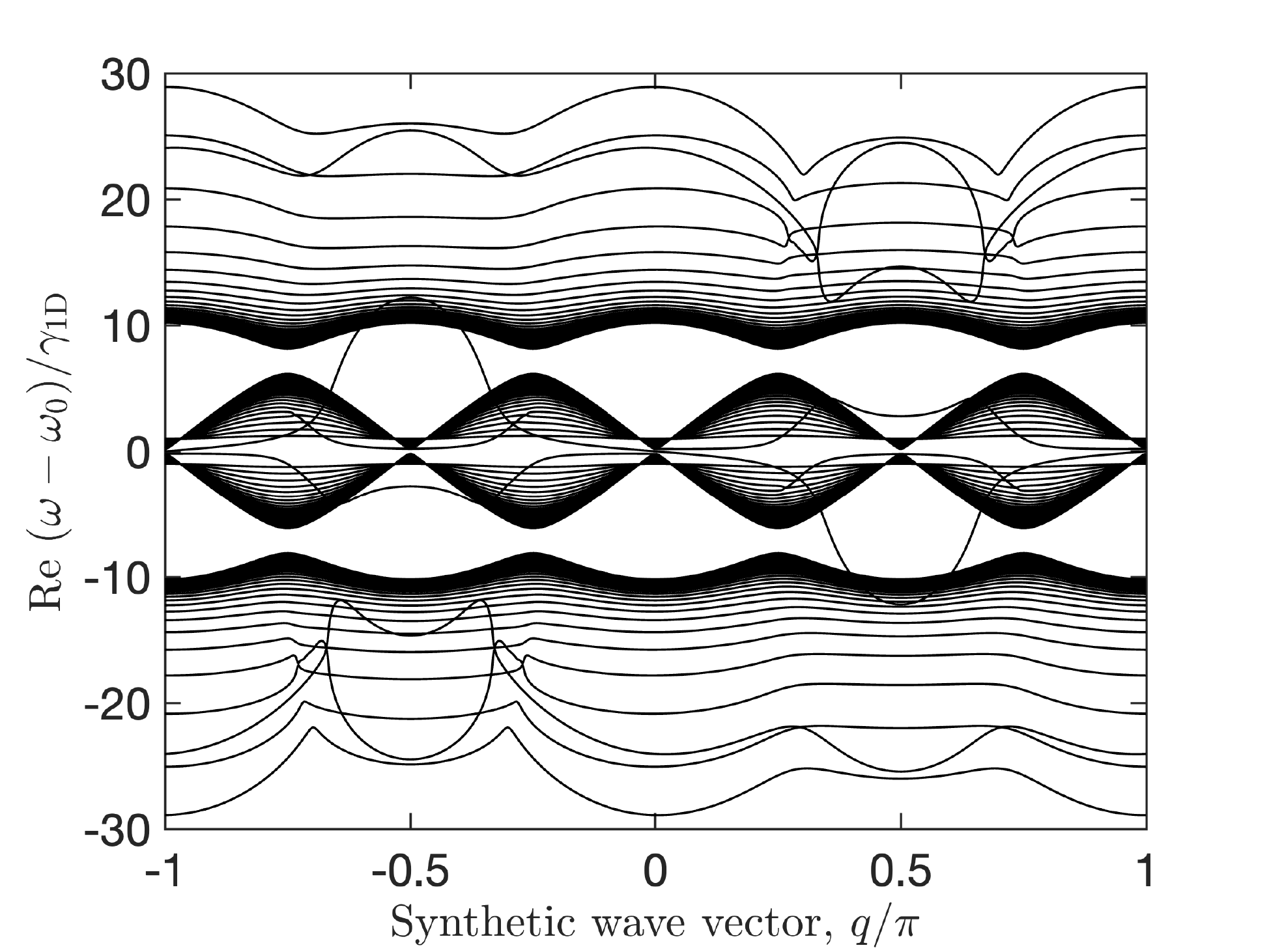}
\caption{Energy spectrum in the ribbon geometry, calculated for the Hamiltonian Eq.~\eqref{eq:ribbon}  
for $N=201$, $a=\Omega=5\gamma_{\rm 1D}$, $\gamma=0$.}\label{fig:5}
\end{figure}

Such preferential excitation of the array edges can be interpreted as a manifestation of the topological edge states in the synthetic magnetic field, induced by the time-dependent resonance frequency modulation~\cite{Lewenstein2014,Fan2016}. 
The reason behind such explanation is that Eqs.~\eqref{eq:hmain} can be formally understood  as a problem in two dimensions spanned by the physical coordinate $j$ (emitter number) and the harmonic number $m$, that played the role of the coordinate in the synthetic dimension.
To illustrate the formation of edge states we first consider an auxiliary eigenvalue problem
\begin{multline}\label{eq:hmain1}
\omega_0p_j^{(m)}+A\qty(\e^{-\rmi \alpha_j}p_j^{(m-1)}+\e^{\rmi \alpha_j}p_j^{(m+1)})\\-\rmi \gamma_{\rm 1D}\sum\limits_{j'=1}^N 
\e^{\rmi\varphi|j-j'|}p_{j'}^{(m)}=\omega p_j^{(m)}\:.
\end{multline}
The main difference between original Eq.~\eqref{eq:hmain} and Eq.~\eqref{eq:hmain1} is that we have neglected the term $m\Omega$. As a result, the problem Eq.~\eqref{eq:hmain1} has become periodic both in the harmonic number $m$ (with the period 1) and in the coordinate $j$ (with the period $4$ for $\alpha_j=j\pi/2$). Next, we use the periodic boundary conditions in the synthetic dimension and seek for the solutions $p_j^{(m)}\propto p_j\e^{\rmi qm}$ with the synthetic wave vector $q$. These are found as eigenstates of the following effective  Hamiltonian
\begin{equation}
\label{eq:ribbon}
H_{jj'}(q)=[\omega_0+2A\cos (\alpha j+q)]\delta_{jj'}-\rmi \gamma_{\rm 1D}\e^{\rmi \varphi|j-j'|}\:.
\end{equation}
with $\alpha=\pi/2$.
The problem Eq.~\eqref{eq:ribbon} is similar to the celebrated Aubry-Andre-Harper (AAH) problem, describing the dynamics of an electron on a two-dimensional square lattice in a transverse magnetic field~\cite{bernevig2013}. The role of magnetic flux in the AAH model is played by the modulation phase gradient $\alpha$. The only difference from the original AAH model is that instead of the nearest-neighbor tight-binding couplings the last term in Eq.~\eqref{eq:ribbon} describes long-range waveguide-mediated couplings.
However, as has been demonstrated in Ref.~\cite{Poshakinskiy2014} for a slightly different setup, the long-range couplings do not affect main features of the Aubry-Andre-Harper model:  nonzero Chern numbers for the allowed bands and formation of topological edge states in the band gaps. In order to illustrate these effects we present in Fig.~\ref{fig:5} numerically calculated eigenfrequencies of Eq.~\eqref{eq:ribbon} in the long  array  with $N=201$ emitters depending on the wave vector $q$ in the synthetic dimension. The calculation clearly demonstrates formation of 4 allowed bands, separated by 3 band gaps with 2 edge states per each band gap. These edge states span the band gaps when the wave vector changes from $-\pi$ to $\pi$ which is a clear indication of their topological origin. The overall band structure is also in full qualitative agreement  with the calculation in Ref.~\cite{Fan2016}, where a tight-binding problem has been considered with the last term in Eq.~\eqref{eq:ribbon} replaced by $t\delta_{|j-j'|,1}$. 

We now turn back from Eqs.~\eqref{eq:hmain1} to our  original system of equations ~\eqref{eq:hmain}. The original system has the potential $m\Omega$, that is equivalent to the constant electric field applied in the synthetic direction. Qualitatively, such electric field should not destroy localization in the physical direction $j$. The main effect of the term $m\Omega$ is just the localization in the synthetic direction $m$, that is very similar to the Wannier-Stark localized states arising for electrons on a 1D lattice in an electric field~\cite{Gluck2002}. Thus, the topological edge states, seen in Fig.~\ref{fig:4}, should persist when the $m\Omega$ term is taken into account, but they will become localized in the synthetic direction of the wave vector $q$. Such interpretation is consistent with our analysis of the $\left<|p_j^{(m)}|^2\right>$ dependence: it is concentrated at small harmonic numbers $m$   (localization in synthetic direction) and at the array edges $j=1,j=N$ (localization due to topological origin of edge states). It also explains the concentration of the occupations at the array edges, manifested in Fig.~\ref{fig:4}(a).

\section{Summary}\label{sec:summary}
To summarize, we have developed a numerical and analytical theory of ratchet effect in arrays of resonant light emitters, coupled to a one-dimensional waveguide. The essence of the effect is the asymmetric spatial distribution of the emitter occupations under symmetrical excitation by electromagnetic wave from both sides of the array. The asymmetry is driven  by the external periodic time modulation of their resonance frequencies. 

We have started by the consideration of the simplest setup of just a pair of coupled emitters. In this case we were able to derive a simple analytical perturbation theory in the coupling between the emitters, that satisfactory describes the results of numerical calculation. We find the optimal conditions for the maximal occupation asymmetry: the pumping should be detuned by a modulation frequency from the emitter resonance and the modulation frequency $\Omega$ and amplitude $A$ should be of the same order, $2A\approx \Omega$.

Next, we have considered larger arrays with up to 20 emitters, with the modulation phase linearly distributed along the array. In this case the modulation makes the structure topologically nontrivial: an effective magnetic field arises in the synthetic two-dimensional space spanned by the physical coordinate of the emitters and the Fourier harmonic number of the time dependence of the emitter polarizations driving a synthetic quantum Hall phase. As a result of formation of topological edge states only the emitters either at the left or right edges of the array are occupied  in a broad range of pump frequencies. This edge localization enhances the asymmetry of the emitter occupation distribution  and the ratchet effects.

The ratchet effects, considered in this work, are essentially classical. It would also be  very interesting to study the many-body ratchet effects in the quantum emitters arrays in the multiple excitation  regime. In this case one can expect one more mechanism of symmetry breaking based on the intrinsic quantum nonlinearity of the emitters~\cite{Fedorov2018}. Our current results can  be readily tested in state-of-the-art platforms of waveguide quantum electrodynamics. We hope that the time modulation approach will provide  novel opportunities to control future quantum chips. 

\begin{acknowledgments}
The authors acknowledge useful discussions with A.V.~Poshakinskiy and E.~S.~Redchenko.
{The numerical calculation of the distribution of emitter polarizations, performed by  A.~N.~P., have been funded by the Russian Science Foundation Grant No.~19-12-00051.}
Development of the analytical theory by L.~E.~G. has been  supported by RFBR--DFG (project 21-52-12015) and Foundation for the Advancement of Theoretical Physics and Mathematics ``BASIS''.
\end{acknowledgments}
%

\end{document}